\documentclass[twocolumn,aps,prl,showpacs]{revtex4}
\usepackage[T1]{fontenc}
\usepackage[latin9]{inputenc}
\usepackage{amsmath}
\usepackage{amssymb}
\usepackage{graphicx}
\usepackage{esint}

\makeatletter
\@ifundefined{textcolor}{}
{%
 \definecolor{BLACK}{gray}{0}
 \definecolor{WHITE}{gray}{1}
 \definecolor{RED}{rgb}{1,0,0}
 \definecolor{GREEN}{rgb}{0,1,0}
 \definecolor{BLUE}{rgb}{0,0,1}
 \definecolor{CYAN}{cmyk}{1,0,0,0}
 \definecolor{MAGENTA}{cmyk}{0,1,0,0}
 \definecolor{YELLOW}{cmyk}{0,0,1,0}
}

\makeatother

\begin{document}

\title{Optical control of a magnetic Feshbach resonance in ultracold Fermi
gases}

\author{Zhengkun Fu$^{1}$, Pengjun Wang$^{1}$, Lianghui Huang$^{1}$, Zengming
Meng$^{1}$, Hui Hu$^{2}$, and Jing Zhang$^{1,\dagger}$}

\affiliation{$^{1}$State Key Laboratory of Quantum Optics and Quantum Optics
Devices, Institute of Opto-Electronics, Shanxi University, Taiyuan
030006, P.R. China\\
 $^{2}$Centre for Atom Optics and Ultrafast Spectroscopy, Swinburne
University of Technology, Melbourne 3122, Australia}

\date{\today}
\begin{abstract}
We use laser light near-resonant with a molecular bound-to-bound transition
to control a magnetic Feshbach resonance in ultracold Fermi gases
of $^{40}$K atoms. The spectrum of excited molecular states is measured
by applying a laser field that couples the ground Feshbach molecular
state to electronically excited molecular states. Nine strong bound-to-bound
resonances are observed below the $^{2}P_{1/2}+^{2}S_{1/2}$ threshold.
We use radio-frequency spectroscopy to characterize the laser-dressed
bound state near a specific bound-to-bound resonance and show clearly
the shift of the magnetic Feshbach resonance using light. The demonstrated
technology could be used to modify interatomic interactions with high
spatial and temporal resolutions in the crossover regime from a Bose-Einstein
condensate (BEC) to a Bardeen-Cooper-Schrieffer (BCS) superfluid.
\end{abstract}


\maketitle The ability to control the strength of interatomic
interactions has led to revolutionary progress in the field of
ultracold atomic gases \cite{Chin2010}. Magnetic-field-induced
Feshbach resonance is one of the such powerful tools and has been
used widely in atomic gases of alkali atoms to understand strong
correlation of quantum many-body systems \cite{Chin2010}. An
alternative technique for tuning interatomic interactions is optical
Feshbach resonance (OFR) \cite{Fedichev1996,Bohn1999}, in which free
atom pairs are coupled to an excited molecular state by laser field
near a photoassociation resonance. OFR is particularly useful for
controlling interatomic interactions in atomic gases of alkali earth
atoms \cite{Enomoto2008,Yamazaki2010,Blatt2011,Yan2013}, because of
the lack of magnetic structure in the ground state of these atoms.
It also offers more flexible control on the interaction strength
with high spatial and temporal resolutions, since the laser
intensity can be varied on short length and time scales. However,
OFR often suffers from rapid loss of atoms due to the light-induced
inelastic collisions between atoms. Recently, optical lasers in
combination with magnetic Feshbach resonances have been developed to
modify the interatomic interaction in a Bose gas
\cite{Bauer2009,Junker2008} atoms and have been shown to reduce the
loss rate by an order of magnitude compared with the ordinary OFR in
$^{87}$Rb \cite{Bauer2009,Bauer2009-PRA}. In this Letter, we
demonstrate the realization of such a laser-controlled magnetic
Feshbach resonance in a strongly interacting Fermi gas of $^{40}$K
atoms.

Strongly interacting atomic Fermi gas is a clean and easily controllable
system with rich and intriguing physical properties \cite{OHara2002,Giorgini2008}.
It provides a new platform for solving some challenging problems in
condensed matter physics and for quantum simulating novel exotic quantum
states of matter, such as the high-temperature superconductivity and
BEC-BCS crossover \cite{Bloch2008}. Magnetic Feshbach resonance has
already been shown to play a key role in exploring strongly interacting
Fermi gases with balanced or imbalanced spin-populations, leading
to the creation of molecules \cite{Regal2003a,Jochim2003,Martin2003,Bourdel2003},
realization of fermionic superfluidity \cite{Regal2004,Martin2004,Martin2005,Zwierlein2006,Partridge2006}
and discovery of fermionic universality \cite{Ho2004,Hu2007,Salomon2010,Horikoshi2010,Ku2012}.
The additional independent control of interatomic interactions with
laser light, as demonstrated in this work, would give rise to the
possibility of revealing new quantum phases of atomic Fermi gases.

To characterize a strongly interacting atomic Fermi gas, radio-frequency
(rf) spectroscopy is a valuable experimental tool. It has been used
extensively to determine the $s$-wave scattering length near a Feshbach
resonance by directly measuring the rf shift induced by mean-field
interactions \cite{Regal2003b}, to demonstrate many-body effects
and quantum unitarity \cite{Gupta2003}, and to probe the occupied
spectral function of single-particle states and the energy dispersion
through BEC-BCS crossover \cite{Stewart2008}.

Here we experimentally investigate magnetic Feshbach resonance in
combination with laser light and characterize the laser-modified bound
state by using rf spectroscopy. Feshbach molecules are created in
a $^{40}$K atomic Fermi gas consisting of an equal mixture of atoms
in the $|9/2,-9/2\rangle\equiv|F=9/2,m_{F}=-9/2\rangle$ and $|9/2,-7/2\rangle$
hyperfine states, by ramping a magnetic bias field from above the
broad $s$-wave Feshbach resonance at $B_{0}=202.20\pm0.02$ G \cite{Gaebler2010}
to certain values below the resonance. We measure the spectrum of
the excited molecular state by applying a laser field near resonant
with transitions between the ground Feshbach molecular state and the
electronically excited molecular states, and show that there are nine
strong bound-to-bound transitions. At a specific bound-to-bound transition,
we demonstrate that the magnetic Feshbach resonance is notably shifted
by a detuned laser light, as evidenced by the magnetic field and frequency
detuning dependence of the laser-dressed bound state in rf spectroscopy.
Our measurements are in good agreement with a simple theoretical model.

\begin{figure}[t]
\begin{centering}
\includegraphics[clip,width=0.4\textwidth]{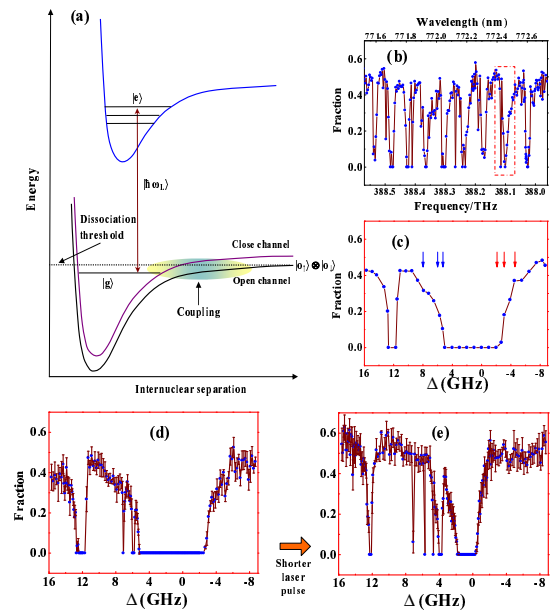}
\par\end{centering}

\caption{(Color online) \textbf{Energy level diagram and excited-state spectroscopy}.
\textbf{(a)} Schematic diagram of the energy levels. A magnetic Feshbach
resonance couples atoms in the incoming open channel $|o_{\uparrow}\rangle\otimes|o_{\downarrow}\rangle$
to a bound dimer state $|g\rangle$ in a different potential (the
closed channel). A laser with frequency $\omega_{L}$, near resonant
with a bound-to-bound transition, transfers molecules in $|g\rangle$
to one of the electronically excited dimer states $|e\rangle$. \textbf{(b)}
Bound-to-bound spectroscopy for $^{40}K_{2}$ molecules below the
$^{2}P_{1/2}+^{2}S_{1/2}$ threshold. \textbf{(c)} Enlarged view of
the 8th strong resonance labeled in (b), plotted as a function of
the detuning $\Delta=(2\pi\hbar)(\omega_{L}-\omega_{eg}$), where
$\omega_{eg}\simeq388103.7$ GHz is the resonance frequency of the
8th bound-to-bound transition. Arrows indicate the detunings used
in Fig. \ref{fig3}(a). \textbf{(d)} Bound-to-bound spectroscopy of
(c) with fine frequency resolution. \textbf{(e)} The spectroscopy
with a shorter laser pulse duration ($1.5$ ms) with the same laser
power, which shows multi-substructure at the vibrational level.}

\label{fig1}
\end{figure}

The experimental procedure starts with a degenerate Fermi gas of about
$N\simeq2\times10^{6}$ $^{40}$K atoms in the $|9/2,9/2\rangle$
internal state, which has been evaporatively cooled to $T/T_{F}\simeq0.3$
with bosonic $^{87}$Rb atoms in the $|F=2,m_{F}=2\rangle$ state
inside a crossed optical trap \cite{Xiong2008,Xiong2010a,Xiong2010b,Wang2011,Wang2012}.
Here $T$ is the temperature and $T_{F}$ is the Fermi temperature
defined by $T_{F}=E_{F}/k_{B}=(6N)^{1/3}\hbar\overline{\omega}/k_{B}$
with a geometric mean trapping frequency $\overline{\omega}$. A $780$
nm laser pulse of $0.03$ ms is used to remove all the $^{87}$Rb
atoms in the mixture without heating $^{40}$K atoms. Subsequently,
fermionic atoms are transferred into the lowest state $|9/2,-9/2\rangle$
via a rapid adiabatic passage induced by a rf field with duration
of $80$ ms at $B\simeq4$ G. To prepare a Fermi gas with equal spin-population
in the $|9/2,-9/2\rangle$ and $|9/2,-7/2\rangle$ states, a homogeneous
magnetic bias field along the $z$-axis (gravity direction) produced
by two coils (operating in the Helmholtz configuration) is raised
to about $B\simeq219.4$ G and then a rf ramp around $47.45$ MHz
is applied for $50$ ms. These two hyperfine states form the incoming
state $|o_{\uparrow}\rangle\otimes|o_{\downarrow}\rangle=|9/2,-9/2\rangle\otimes|9/2,-7/2\rangle$
in the open channel for a pair of atoms, as shown in Fig. \ref{fig1}(a).
We use a magnetically controlled Feshbach resonance at $B_{0}=202.20\pm0.02$
G to convert a pair of atoms into an extremely weakly bound molecules.
With an adiabatic magnetic field sweep across the resonance, molecules
are prepared in the electronic ground dimer sate $|g\rangle$ in the
closed channel.

To optically control the magnetic Feshbach resonance, a laser beam
is derived from a Ti:Sapphire laser using an acousto-optical modulator
in the single-pass configuration, which allows precise control of
the laser intensity and duration time of the pulse. The laser is implemented
as a traveling wave with a $1/e^{2}$ radii of $200$ $\mu m$ and
its linear polarization is perpendicular to the magnetic bias field
of Feshbach resonance. We apply the laser pulse in a rectangular temporal
shape with intensity $I$ up to $15$ ms. Near resonant with one of
the $|g\rangle-|e\rangle$ bound-to-bound transitions, the laser induces
loss in the population of Feshbach molecules due to the excitation
to $|e\rangle$ and subsequent fast spontaneous radiative decay to
unobserved states. In order to determine the number of remaining molecules
in the trap, after turning off the laser, a gaussian-shape rf pulse
with duration about 400 $\mu s$ is applied to dissociate the remaining
molecules into free atoms in the state $|9/2,-9/2\rangle\otimes|9/2,-5/2\rangle$.
After the rf pulse, we abruptly turn off the optical trap and magnetic
field, and let the atoms ballistically expand for $12$ ms in a magnetic
field gradient applied along the $\hat{z}$ axis and then take absorption
image along the $\hat{y}$ direction. The atoms in different hyperfine
states $N_{\sigma}$ ($\sigma=|-7/2\rangle,|-5/2\rangle...$) are
spatially separated and analyzed, from which we determine the fraction
$N_{-5/2}/(N_{-5/2}+N_{-7/2})$ for different laser frequencies to
obtain the the spectrum of the excited molecular states.

\begin{figure}[t]
\begin{centering}
\includegraphics[clip,width=0.35\textwidth]{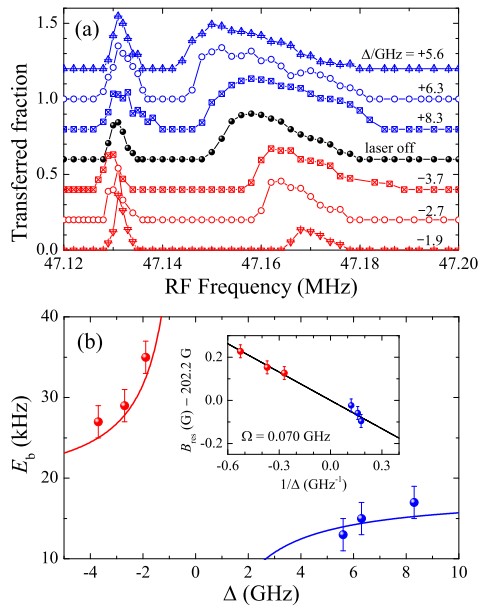}
\par\end{centering}

\caption{(Color online) \textbf{rf spectroscopy and binding energy of the laser-dressed
bound state near the Feshbach resonance} \textbf{$B_{0}=202.20\pm0.02$
G}. With the light off, Feshbach molecules are created below the resonance
at $B=201.60$ G, corresponding to an $s$-wave scattering length
$a_{s}\simeq2216a_{0}$, where $a_{0}$ is the Bohr radius. The dimensionless
interaction parameter of the Fermi cloud is $1/(k_{F}a_{s})\simeq0.62$.
\textbf{(a)} The rf spectroscopy at different detunings, off-set vertically
for clarity. \textbf{(b)} The binding energy as a function of the
detuning. The solid lines are the theoretical predictions from a simple
theory as outlined in the supplemental materials. The inset shows
the resonance position of the shifted Feshbach resonance as a function
of the inverse detuning.}

\label{fig2}
\end{figure}

Fig. \ref{fig1}(b) reports the bound-to-bound spectroscopy for excited
$^{40}K_{2}$ molecules below the $^{2}P_{1/2}+^{2}S_{1/2}$ threshold
at the magnetic field $B=201.60$ G. Here the bound molecules are
illuminated with the laser intensity of $I=50$ mW and the pulse duration
time of $15$ ms. The frequency of the rf pulse used to dissociate
molecules is fixed to a value that is about $20$ kHz larger than
the Zeeman splitting between the hyperfine states $|9/2,-7/2\rangle$
and $|9/2,-5/2\rangle$, which corresponds to the transition from
the bound molecules to the free atom state $|9/2,-9/2\rangle\otimes|9/2,-5/2\rangle$.
The spectrum in Fig. \ref{fig1}(b) covers a wavelength range from
$771.5$ nm to $772.7$ nm with coarse frequency resolution about
$3$ GHz. We find nine strong bound-to-bound loss resonances with
an average frequency spacing of $\sim64$ GHz. A specific loss resonance
at $\omega_{L}\simeq388103.7$ GHz is highlighted in Fig. \ref{fig1}(c)
with an enlarged view. We have extended the frequency scan from $770.5$
nm to $780.0$ nm, but can not detect any further strong bound-to-bound
resonances. According to the photoassociation data for $^{39}K_{2}$
\cite{Stwalley1999}, there are three long range attractive potentials
(labelled by the quantum numbers $0_{u}^{+}$, $0_{g}^{-}$ and $1_{g}$
based on the Hund's case (c) selection rules \cite{Stwalley1999})
adiabatically connected to the $^{2}P_{1/2}+^{2}S_{1/2}$ threshold,
as a result of spin-obit interaction \cite{Jones2006}. The observed
resonances should be related to the bound states of these three potentials.
In principle, each observed bound states would display rich substructures
induced by vibration, rotation, hyperfine interaction, and Zeeman
interaction of molecules. As shown in Fig. \ref{fig1} (d) and (e),
we are able to resolve some of these substructures by increasing the
laser frequency resolution or reducing the pulse duration. Fig. \ref{fig1}(e)
reports the spectroscopy near the $\omega_{L}\simeq388103.7$ GHz
resonance with the pulse duration of $1.5$ ms and the frequency resolution
of $100$ MHz. Multi-substructures at the vibrational level can be
clearly identified. A careful parameterization of these fine structures
will be addressed in the future, as it requires a much better precision
of the frequency calibration.

\begin{figure}[t]
\begin{centering}
\includegraphics[clip,width=0.48\textwidth]{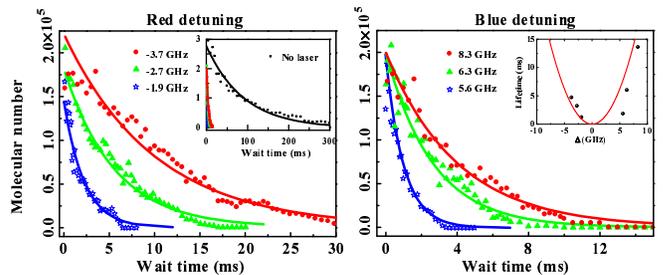}
\par\end{centering}

\caption{(Color online) \textbf{Decay of the number of ground-state bound molecules
at different laser detuning}s: red detuning (a) and blue detuning
(b). The inset in (a) shows the decay without laser. The inset in
(b) reports the lifetime vs. detuning, from which we extract $\gamma\simeq2\pi\times6$
MHz by using a parabolic fit.}

\label{fig3}
\end{figure}

Now we use the above bound-to-bound spectroscopic information to shift
the magnetic Feshbach resonance. We focus on the strong resonance
at $\omega_{eg}\simeq388103.7$ GHz. To keep the rate for incoherent
processes as low as possible, we increase the detuning and intensity
of the laser light. For a large detuning, the excited molecular state
is essentially empty and may therefore be adiabatically eliminated.
The resulting dressed ground molecular state acquires an a.c.-Stark
shift (see the supplemental materials for details),
\begin{equation}
\delta=\frac{\Omega^{2}}{4\left(\Delta+i\gamma/2\right)}\simeq\frac{\Omega^{2}}{4\Delta}-\left(\frac{\Omega^{2}}{4\Delta^{2}}\right)\frac{i\gamma}{2},
\end{equation}
where $\Omega$ is the Rabi frequency of laser beam,
$\Delta=(2\pi\hbar)(\omega_{L}-\omega_{eg})$ is the detuning, and
$\gamma\sim2\pi\times6$ MHz stands for the fast spontaneous
radiative decay of the excited molecular state \cite{Bauer2009-PRA}.
Our measurements are performed under the condition
$\Omega\ll\Delta\sim(2\pi\hbar)\times1$ GHz, so that the effective
decay rate
$\gamma_{eff}\equiv(\gamma\Omega^{2}/8\Delta^{2})\sim2\pi\times1$
kHz and therefore the atomic loss should be greatly suppressed. Such
a suppression was also observed in the recent experiment for bosonic
$^{87}$Rb atoms \cite{Bauer2009-PRA}, where a large detuning was
used.

Our experimental sequence is almost the same as before, except that
the light intensity is ramped within $15$ ms to a final power of
$\sim85$ mW. A rf field is then applied to dissociate molecules into
the free-atom state. After the rf pulse, the optical trap, magnetic
field and laser light are abruptly turned off. The relative spin
population of the final state $|9/2,-5/2\rangle$, the fraction
$N_{-5/2}/(N_{-5/2}+N_{-7/2})$, is then measured as a function of
the rf frequency. Fig. \ref{fig2}(a) shows the rf spectroscopy at
different detunings below the Feshbach resonance (at the magnetic
field $B=201.60$ G). The corresponding binding energy $E_{b}$ of the
ground-state dressed molecules, extracted following the procedure in
the supplemental materials, is given in Fig. \ref{fig2}(b). Compared
with the data in the absence of laser light, the red and blue
detunings of the light tend to increase and decrease the binding
energy, respectively. The smaller $\left|\Delta\right|$, the larger
change in the binding energy. Figs. \ref{fig3}(a) and \ref{fig3}(b)
report the population of the ground-state bound molecules as a
function of the duration time of the laser pulse for different laser
detunings used in Fig. \ref{fig2}(a). With the light on, the loss
rate of ground-state molecules increases rapidly with decreasing the
absolute value of detuning. The observed lifetime is typically about
several milliseconds, in agreement with the estimated effective
decay rate $\gamma_{eff}\sim2\pi\times1$ kHz. It is encouraging that
the longest lifetime in our measurements (at the maximum blue
detuning) can reach $\sim10$ ms, which is only 1 order of magnitude
lower than that without laser. Thus, our optically controllable
Fermi system turns out to be very stable.

We develop a simple two-body theory to understand the observed data.
As shown in the supplemental materials, we present both the approximate
analytic result and full numerical calculation for the binding energy.
Near the broad Feshbach resonance of $^{40}$K atoms, to a good approximation
we have $\sqrt{E_{b}}\propto B-B_{res}(\Omega)$ or
\begin{equation}
\sqrt{\frac{E_{b}}{E_{b}\left(\Omega=0\right)}}=\frac{B-B_{res}\left(\Omega\right)}{B-B_{0}},\label{eq:Eb}
\end{equation}
where $B_{res}(\Omega)\simeq B_{0}-\Omega^{2}/(4\Delta\mu_{ag})$
is the light-shifted resonance field and $\mu_{ag}\equiv2\mu_{a}-\mu_{g}\simeq2\mu_{B}$
($\mu_{B}$ is the Bohr magneton) is the difference in the magnetic
moments of atoms and molecules. Using Eq. (\ref{eq:Eb}), we extract
$B_{res}(\Omega)$ from the binding energy and report it in the inset
of Fig. \ref{fig2}(b). A linear fit of $1/[B_{res}(\Omega)-B_{0}]$
with respect to $\Delta$ leads to the determination of the Rabi frequency
\cite{footnote}, $\Omega=(2\pi\hbar)\times(0.070\pm0.007)$ GHz,
which is much smaller than the absolute magnitude of the detuning,
as we may anticipate. We have used $\Omega=(2\pi\hbar)\times0.070$
GHz to numerically predict the binding energy without any approximation.
The results, shown by the solid lines in the main figure of Fig. \ref{fig2}(b),
agree well with the data. Thus, we conclude that the light-shifted
Feshbach resonance is well understood in the regime of large detunings.

\begin{figure}[t]
\begin{centering}
\includegraphics[clip,width=0.4\textwidth]{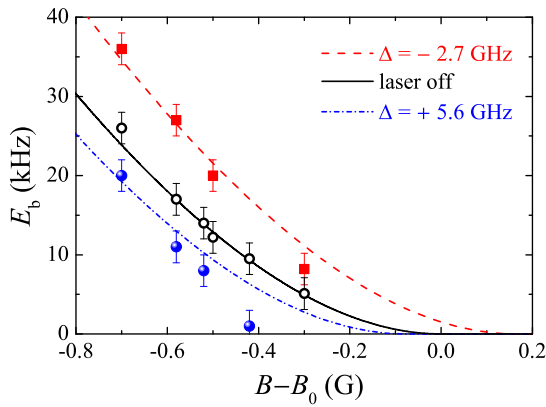}
\par\end{centering}

\caption{(Color online) \textbf{Binding energy of the laser-dressed bound state
at different detunings. }The data at the red and blue detunings (symbols),
$\Delta=(2\pi\hbar)\times(-2.7)$ GHz and $\Delta=(2\pi\hbar)\times(+5.6)$
GHz, have been compared with the two-body predictions (lines). In
the numerical calculation, we take $\Omega=(2\pi\hbar)\times0.070$
GHz.}

\label{fig4}
\end{figure}

\begin{figure}[t]
\begin{centering}
\includegraphics[clip,width=0.4\textwidth]{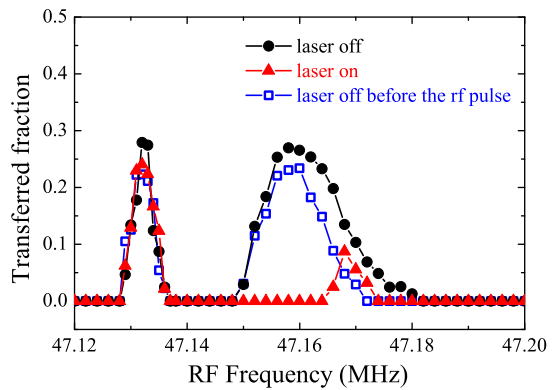}
\par\end{centering}

\caption{(Color online) \textbf{Demonstration of the high temporal resolution
of the laser-dressed bound state at} \textbf{$B_{0}=201.60$ G}. Solid
circles: rf spectroscopy with the laser off. Solid triangles: rf spectroscopy
with the laser on. Empty squares: rf spectroscopy with the laser light
abruptly turned off just before the switch-on of the rf field. The
laser detuning is $\Delta=(2\pi\hbar)\times(-1.9)$ GHz.}

\label{fig5}
\end{figure}

We also measure the rf spectroscopy and determine the binding energy
at different magnetic fields while fixing the laser detuning, as reported
in Fig. \ref{fig4}. In this scenario, it is clear that, to prepare
a Fermi cloud with the same interatomic interaction (i.e., to have
the same binding energy in Fig. \ref{fig4}), one can either tune
the magnetic field without the light, or tune the laser detuning at
a given $B$ field. The latter manner is ideally suited to explore
the non-equilibrium dynamics or equilibrium novel states of matter
of strongly interacting Fermi gases, as the laser light can be varied
over short time and length scales. In Fig. \ref{fig5}, we demonstrate
the high temporal resolution of the optically controlled Feshbach
resonance. When the laser light is abruptly turned off just before
the rf field is applied, the rf spectroscopy is roughly the same as
that without the light, in sharp contrast with the result for which
the laser is always on. This indicates that the optical control of
the Feshbach resonance is almost instantaneous and is effective within
a time scale less than $400$ $\mu s$ (i.e., the duration of the
rf pulse).

In conclusions, we have demonstrated the technology of controlling
the magnetic Feshbach resonance using laser light in ultracold atomic
Fermi gases. The spectrum of electronically excited molecular states
of $^{40}K_{2}$ below the $^{2}P_{1/2}+^{2}S_{1/2}$ threshold has
been measured by applying a laser field near resonant with the bound-to-bound
transitions between the ground and excited molecular states. Nine
strong bound-to-bound resonances have been identified below the $^{2}P_{1/2}+^{2}S_{1/2}$
threshold, each of which can be easily used to control the magnetic
Feshbach resonance of $^{40}$K atoms. We have characterized the laser-induced
shift of the Feshbach resonance in the large detuning regime and the
stability of the system, using the standard tool of radio-frequency
spectroscopy, and have understood the data within a simple two-body
theory. The optical tunability of interatomic interactions, as demonstrated
in this work, paves a new way to explore the fascinating quantum many-body
system of strongly interacting Fermi gases.
\begin{acknowledgments}
We thank Dajun Wang for useful discussions. This research is supported
by the National Basic Research Program of China (Grant No. 2011CB921601),
NSFC (Grant No. 11234008), NSFC Project for Excellent Research Team
(Grant No. 61121064), and Doctoral Program Foundation of the Ministry
of Education China (Grant No. 20111401130001). H.H. is supported by
the ARC Discovery Project (Grant No. DP0984522).

$^{\dagger}$Correspondence should be addressed to Jing Zhang
(jzhang74@aliyun.com, jzhang74@sxu.edu.cn).\end{acknowledgments}


\begin{thebibliography}{10}
\bibitem{footnote}Experimentally, the energy splitting $h\omega_{eg}$
can not be precisely determined in the first place. In the linear
fitting, thus we have used the expression $1/[B_{res}(\Omega)-B_{0}]=-(4\mu_{ag}/\Omega^{2})(\Delta+\delta)$
to allow a correction $\delta$ to the pre-determined value of $\omega_{eg}$.
The number $\omega_{eg}\simeq388103.7$ GHz that we quoted earlier
already takes into account the correction $\delta$. From the linear
fitting, we estimate approximately $\omega_{eg}\simeq388103.7\pm0.5$
GHz.
\end{thebibliography}
\end{document}